\documentclass[pdftex,twocolumn,epjc3]{svjour3}          % twocolumn

\RequirePackage[T1]{fontenc}

\smartqed  % flush right qed marks, e.g. at end of proof

\RequirePackage{mathptmx}      % use Times fonts if available on your TeX system
\RequirePackage{flushend}
\RequirePackage[numbers,sort&compress]{natbib}
\RequirePackage[colorlinks,citecolor=blue,urlcolor=blue,linkcolor=blue]{hyperref}

\usepackage{latexsym}
\usepackage{amsmath}
\usepackage{amssymb}
\usepackage{latexsym}
\usepackage{amsfonts}
\usepackage{graphicx}
\usepackage{hyperref}

\journalname{Eur. Phys. J. Plus}

\begin{document}

\title{Reconstruction of extended inflationary potentials for attractors}
\author{\href{https://orcid.org/0000-0003-3797-4370}{Qing Gao}\thanksref{e1,addr1}
        \and
       \href{https://orcid.org/0000-0001-5065-2259}{Yungui Gong}\thanksref{e2,addr2}
}

\thankstext{e1}{e-mail: gaoqing1024@swu.edu.cn}
\thankstext{e2}{e-mail: yggong@hust.edu.cn (corresponding author) }

\institute{School of Physical Science and Technology, Southwest University, Chongqing 400715, China \label{addr1}
\and
School of Physics, Huazhong University of Science and Technology, Wuhan, Hubei 430074, China \label{addr2}
}

\date{Received: 27 April 2018 / Revised: 14 October 2018}

\maketitle

\begin{abstract}
We give the procedure to reconstruct the extended inflationary potentials for attractors and use the $\alpha$ attractor
and the constant-roll model as examples to show how to
reconstruct the class of extended inflationary potentials in the strong coupling limit.
We also derive the strong coupling condition which the coupling constant $\xi$ satisfies.
The class of extended inflationary potentials has the same attractor,
and the reconstructed extended inflationary potentials are consistent with the observational constraints.
The existence of the attractors from extended inflationary models
further challenges the model discrimination just by the observables $n_s$ and $r$.
\end{abstract} %end of abstract

\section{Introduction}

The T model with the potential $V(\phi)\sim \tanh^{2n}(\phi/\sqrt{6})$ \cite{Kallosh:2013hoa},
the E model with the potential $V(\phi)\sim [1-\exp(-\sqrt{2/3}\,\phi)]^n$ \cite{Kallosh:2013maa},
the Higgs inflation with the nonminimal coupling $\xi\psi^2 R$ in the
strong coupling limit $\xi\gg 1$ \cite{Kaiser:1994vs,Bezrukov:2007ep},
and the Starobinsky model $R+R^2$ \cite{Starobinsky:1980te} all predict that the scalar spectral tilt $n_s=1-2/N$
and the tensor to scalar ratio $r=12/N^2$ with $N=60$ which are consistent with the Planck results $n_s=0.9645\pm 0.0049$
and $r_{0.002}<0.10$ (95\% C.L.) \cite{Ade:2015lrj}, here $N$ is the number of $e$-folds before
the end of inflation at the horizon exit.
The almost scale invariant power spectrum suggests that the slow-roll parameters $\epsilon$
and $\eta$ are small at the horizon exit, while the duration of inflation is long enough
to solve the problems in the standard big bang cosmology. So it was suspected
that there are some relations among the scalar spectral tilt $n_s-1$, the amplitude of the
scalar perturbation $A_s$ and $N$ \cite{Huang:2007qz,Gobbetti:2015cya}.
Apparently, the mentioned models above support the inverse relation between $n_s-1$ and $N$.
Because the flatness of the potential is characterized by the slow-roll parameters
and the observables are simple functions of the slow-roll parameters to the first
order of approximation, we can reconstruct inflationary potentials by parameterizing
the observables with $N$ \cite{Lin:2015fqa}. Note that the reconstruction can be thought of as the reverse process of finding
$n_s$ and $r$ in terms of $N$ for a given potential, and is valid
for single field slow-roll inflation only. Since higher order corrections are assumed
to be small \cite{Gong:2014cqa}, the effects of higher order slow-roll parameters are not considered,
but this does not mean that higher order slow-roll parameters are exactly zero.
Of course, this reconstruction procedure provides us the information about the
behaviour of the potential during inflation only, it may not work for all potentials \cite{Garcia-Bellido:2014gna}
and may not be the good way to probe
the physics at the early Universe \cite{Martin:2016iqo}.

The reconstruction of inflationary potentials from the primordial scalar spectrum
in the case of scalar field models minimally coupled to gravity was first made
by Hodges and Blumenthal in \cite{Hodges:1990bf}.
By parametrizing the deviation of the equation of state
(equivalent to the slow-roll parameter $\epsilon$) of the inflaton from the cosmological constant with the
inverse power law $\beta/(N+1)^\alpha$, Mukhanov reconstructed a class of inflationary potential in \cite{Mukhanov:2013tua}.
The parameters $\alpha$ and $\beta$ in the Mukhanov parametrization
were fitted to the observational data in \cite{Barranco:2014ira}, and the field
excursion of the inflaton in these models was discussed in \cite{Garcia-Bellido:2014eva,Garcia-Bellido:2014wfa,Boubekeur:2014xva}.
By assuming that the slow-roll
parameters $\epsilon$ and $\eta$ both scale as $1/N^{p}$ at the leading order in the large $N$ limit,
Roest divided some inflationary models into two universal classes with
different behaviors of the scalar to tensor ratio $r$ \cite{Roest:2013fha}.
The classifications of inflationary models were then generalized to include the parametrizations
of $\epsilon(N)$ with the constant, exponential and logarithmic classes \cite{Garcia-Bellido:2014gna},
and to tachyon inflation \cite{Barbosa-Cendejas:2015rba,Fei:2017fub}.
The inverse relation $n_s-1=-\alpha/N$ can be used to derive $r$ \cite{Creminelli:2014nqa},
and the detailed reconstruction of inflationary potentials from the inverse parametrization
was discussed in \cite{Chiba:2015zpa,Lin:2015fqa}.

The Starobinsky model is the simplest nonlinear gravitational theory $f(R)$ \cite{Buchdahl:1983zz}
which is one of the alternative theories of gravity, the other alternative theories of gravity include the scalar-tensor theories of gravity.
It is well known that $f(R)$ gravity can be written as the scalar-tensor theory of gravity \cite{OHanlon:1972xqa,Teyssandier:1983zz},
and through the conformal transformation, the scalar-tensor theory of gravity can be written as the
canonical scalar field minimally coupled to gravity.
After the conformal transformation, the potentials for the Higgs inflation with the nonminimal coupling and the Starobinsky model
are both approximated by the potential $V(\phi)=V_0[1-c\exp(-\sqrt{2/3}\,\phi)]$ in the Einstein frame.
The potentials for the T model and E model have the same approximate behaviors during inflation,
so they all give the same results on $n_s$ and $r$. For the conformal coupling $\xi\psi^2 R$ with $\xi=-1/6$, the fields $\psi$ and $\phi$ in different frames are related as $\phi=\sqrt{6}\tanh^{-1}(\psi/\sqrt{6})$, so the T model potential in the Einstein frame can be obtained from
the monomial potential $\psi^n$ in the Jordan frame \cite{Kallosh:2013hoa,Kallosh:2013maa}.

The same $n_s=1-2/N$ and $r=12/N^2$ can be obtained
from more general potentials $V_J(\psi)=\lambda^2 f^2(\psi)$ with the nonminimal coupling $\xi f(\psi)R$ for
arbitrary functions $f(\psi)$ in the strong coupling limit \cite{Kallosh:2013tua}. If we take $f(\psi)=\psi^2$,
then we recover the Higgs inflation with the nonminimal coupling. Therefore, the $\xi$-attractors generalize the idea
of the Higgs inflation with the nonminimal coupling.
This idea of obtaining the attractor $n_s=1-2/N$ and $r=12/N^2$ by the nonminimal coupling $\xi f(\psi)R$
was then generalized to any attractor we want by more general nonminimal coupling \cite{Yi:2016jqr}.
Starting from different models with the potentials $V^i_J(\psi)$ and nonminimal couplings $\Omega^i(\psi)$ in Jordan frame,
in general we get different potentials $V^i(\phi)=V^i_J(\psi)/\Omega^i(\psi)$ in Einstein frame after the conformal transformation.
However, if we take the strong coupling limit which will be elucidated in the following section,
the universal relation $\phi=\sqrt{3/2}\,\ln\Omega^i(\psi)$ between the fields $\psi$ and $\phi$ is obtained, and
we can choose the potentials $V^i_J(\psi)$ so that $V^i(\phi)$ become the same potential $V(\phi)$ in the strong coupling limit,
and these different models labelled by superscript $i$ have the same $n_s$ and $r$ in the strong coupling limit,
i.e., the $\xi$-attractor is obtained.
In other words, different scalar-tensor theories of gravity may give
the same observables $n_s$ and $r$ in the strong coupling limit if we choose the potentials and nonminial couplings appropriately.
In general, it is possible to obtain any attractor from general scalar-tensor theories of gravity \cite{Galante:2014ifa, Yi:2016jqr}
or $f(R)$ gravity \cite{Nojiri:2010wj,Odintsov:2016vzz,Yi:2016jqr,Choudhury:2017cos}.
By varying the K\"{a}hler curvature, we can derive the $\alpha$ attractor with $n_s=1-2/N$ and $r=12\alpha/N^2$ \cite{Kallosh:2013yoa},
and the $\alpha$ attractor can also be derived from general scalar-tensor theory of gravity \cite{Galante:2014ifa}.
The constant-roll inflation with the slow-roll parameter $\eta_H$ being a constant \cite{Motohashi:2014ppa,Motohashi:2017aob}
breaks the slow-roll condition if $\eta_H$ is not small and has richer physics than slow-roll inflation
because it includes both slow-roll and ultra slow-roll cases \cite{Tsamis:2003px,Kinney:2005vj,Martin:2012pe,Namjoo:2012aa,Yi:2017mxs,Motohashi:2017vdc,Oikonomou:2017bjx,Odintsov:2017qpp,Nojiri:2017qvx,Gao:2017owg,Dimopoulos:2017ged,Ito:2017bnn,
Karam:2017rpw,Cicciarella:2017nls,Anguelova:2017djf,Gao:2018tdb,Gao:2018cpp,Morse:2018kda,Pattison:2018bct}. Furthermore, the ultra slow-roll inflation can
generate large density perturbation to seed the formation of primordial black holes \cite{Kannike:2017bxn,Motohashi:2017kbs,Germani:2017bcs,Gong:2017qlj}.
It is interesting to study the reconstruction of the $\alpha$ attractor and the constant-roll inflation.

In this paper, we combine the idea of reconstruction and obtaining $\xi$ attractors from general scalar-tensor theories of gravity
to reconstruct the extended inflationary potentials \footnote{Extended inflation was first proposed in \cite{La:1989za} by using the Brans-Dicke theory.}.
The paper is organized as follows. In sec. 2,
we give the general formula and procedure for the reconstruction. In sec. 3, we discuss the
reconstruction for the $\alpha$ attractors, and the reconstruction for exponential parametrization is discussed in sec. 4.
We conclude the paper in sec. 5.

\section{The Reconstruction method}

For the single field inflation, to the first order of slow-roll approximation, we have the relation,
\begin{equation}
\label{nsapproxeq3}
\frac{d\ln\epsilon}{dN}=2\eta-4\epsilon,
\end{equation}
so the scalar spectral tilt can be expressed as
\begin{equation}
\label{nsapproxeq4}
n_s-1=-2\epsilon+\frac{d\ln\epsilon}{dN},
\end{equation}
where the slow-roll parameters\footnote{There are other definitions for the slow-roll parameters \cite{Liddle:1994dx,Schwarz:2001vv}. For example,
the Hubble flow slow-roll parameter $\eta_H$ is defined as $\eta_H=-\ddot{\phi}/(H\dot\phi)$
which is related with $\eta$ as $\eta_H\approx \eta-\epsilon$ in
the slow-roll approximation \cite{Liddle:1994dx}. The constant-roll inflation is usually defined as $\eta_H$ being a constant in the literature.}
\begin{equation}
\label{slrpareq1}
\epsilon=\frac{1}{2}\left(\frac{dV/d\phi}{V(\phi)}\right)^2,\quad \eta=\frac{d^2V/d\phi^2}{V}.
\end{equation}
Note that we take the reduced Planck mass $M^2_{pl}=1/(8\pi G)=1$.
Because
\begin{equation}
\label{dphidn1}
d\phi=\frac{dV/d\phi}{V}dN=\mp \sqrt{2\epsilon}dN,
\end{equation}
so
\begin{equation}
\label{nphieq2}
\phi-\phi_e=\pm \int_0^N \sqrt{2\epsilon(N)}dN,
\end{equation}
where the sign $\pm$ depends on the sign of the first derivative of the potential and the scalar
field is normalized by the reduced Planck mass $M_{pl}=1$. Substituting relation \eqref{dphidn1} into eqs.
\eqref{nsapproxeq4} and \eqref{slrpareq1}, we get \cite{Chiba:2015zpa,Lin:2015fqa}
\begin{equation}
\label{sclreq2}
\epsilon=\frac{1}{2}\frac{dV/d\phi}{V}\frac{d\phi}{dN}=\frac{1}{2}\frac{d\ln V}{dN}=\frac{1}{2}(\ln V)_{,N}>0,
\end{equation}
\begin{equation}
\label{nsapproxeq6}
n_s-1= -(\ln V)_{,N}+\left(\ln\frac{V_{,N}}{V}\right)_{,N}=\left(\ln\frac{V_{,N}}{V^2}\right)_{,N}.
\end{equation}
If we parametrize one of the functions as $\epsilon(N)$, $n_s(N)$, $\phi(N)$ or $V(N)$, then we can derive the other
functions by using eqs. (\ref{nsapproxeq4}), \eqref{dphidn1} and (\ref{sclreq2}). So once one of the functions
is parametrized,  in principle we can get the observables
$n_s$ and $r$, and reconstruct the potential $V(\phi)$
by using the relations (\ref{nsapproxeq4}), \eqref{dphidn1}, (\ref{sclreq2}) and (\ref{nsapproxeq6}).
Similarly, we can use the so called $\beta$-function formalism \cite{Binetruy:2014zya,Pieroni:2015cma,Binetruy:2016hna,Cicciarella:2016dnv}
\begin{equation}
\label{betaeq1}
\beta(\phi)=-\frac{2d\ln W(\phi)}{d\phi}\approx \pm \sqrt{2\epsilon},
\end{equation}
to reconstruct the potential, where the so called superpotential $W(\phi)=-2H(\phi)$, and the approximation means the first
order of slow-roll approximation.

Since the adiabatic perturbations are invariant under conformal transformation \cite{Makino:1991sg,Fakir:1992cg,Weenink:2010rr,Kubota:2011re,Gong:2011qe,Prokopec:2013zya,Chiba:2013mha,Postma:2014vaa,Li:2015hga},
the relationships between observable $n_s$ and $r$ in different frames are easily established
although the physical equivalence between Einstein and Jordan frames is unclear \cite{Jarv:2014hma,Kuusk:2016rso,Catena:2006bd,Burns:2016ric,Jarv:2016sow}.
To reconstruct the potential $V_J[\psi(\phi)]$ in the Jordan frame,
we need to have the relation between the potentials in different frames.
For a general scalar-tensor theory in the Jordan frame
\begin{equation}
\label{jscten}
\begin{split}
S=\int d^4x\sqrt{-\tilde{g}}\left[\frac{1}{2}\Omega(\psi)\tilde{R}(\tilde{g})-{1\over
2}\omega(\psi)\tilde{g}^{\mu\nu}
\nabla_{\mu}\psi\nabla_{\nu}\psi \right.\\
\left.-V_J(\psi)\right],
\end{split}
\end{equation}
under the conformal transformations,
\begin{gather}
\label{conftransf1}
g_{\mu\nu}=\Omega(\psi){\tilde g}_{\mu\nu},\\
\label{conftransf2}
d\phi^2=\left[\frac{3}{2}\frac{(d\Omega/d\psi)^2}{\Omega^2(\psi)}+\frac{\omega(\psi)}{\Omega(\psi)}\right]d\psi^2,
\end{gather}
we get the usual Einstein-Hilbet action with a minimally coupled scalar field,
\begin{equation}
\label{escten}
S=\int d^4x\sqrt{-g}\left[\frac{1}{2}R(g)-\frac{1}{2}g^{\mu\nu}\nabla_\mu\phi
\nabla_\nu\phi-V(\phi)\right],
\end{equation}
where $V(\phi)=V_J(\psi)/\Omega^2(\psi)$. Therefore, once we reconstruct the potential $V(\phi)$ with
the relations (\ref{nsapproxeq4}), \eqref{dphidn1} and (\ref{sclreq2}) by parameterizing the observables,
then the corresponding potential for the general scalar-tensor theory in Jordan frame is $V_J(\psi)=\Omega^2(\psi)V[\phi(\psi)]$.
Of course, the models with the potentials $V_J(\psi)=\Omega^2(\psi)V[\phi(\psi)]$ and the conformal factor $\Omega(\psi)$ in Jordan frame transform to the same model with the potential $V(\phi)$
in Einstein frame even though $\Omega(\psi)$ is arbitrary, i.e., these models are degenerate
and they are the same model. However, if we start with a potential $V_J(\psi)$ and an arbitrary conformal factor $\Omega(\psi)$ in Jordan frame,
for different choices of $\Omega(\psi)$ in general we will get different potentials $V(\phi)$ in Einstein frame,
and the derived observables $n_s$ and $r$ will depend on the particular form of $\Omega(\psi)$ and the coupling constant for the nonminimal coupling.

In practice, to reconstruct the potential $V_J(\psi)$, we need to specify the conformal factor $\Omega(\psi)$
and solve eq. \eqref{conftransf2} to get the relationship between $\psi$ and $\phi$. In general, it is difficult
to obtain an analytical relation. However, if the contribution from $\omega(\psi)$ is negligible,  i.e., if the conformal factor satisfies the condition
\begin{equation}
\label{strongcoup1}
\Omega(\psi)\ll \frac{3(d\Omega(\psi)/d\psi)^2}{2\omega(\psi)},
\end{equation}
then we get
\begin{equation}
\label{psiphirel1}
\phi\approx \sqrt{\frac{3}{2}}\ln\Omega(\psi),\quad \Omega(\psi)\approx e^{\sqrt{2/3}\,\phi}.
\end{equation}
Note that the with the above relation \eqref{psiphirel1}, the scalar-tensor theory considered is equivalent
to the $f(R)=\Omega R -2V_J$ gravity with the following relation between $R$ and $\psi$
\begin{equation}
\label{stfreq1}
\Omega(\psi)=\frac{df(R)}{dR},\quad R\frac{d\Omega}{d\psi}=2\frac{dV_J}{d\psi}.
\end{equation}
The condition \eqref{strongcoup1} is called the strong coupling limit.
In the strong coupling limit, we get
\begin{equation}
\label{pontrel1}
V_J(\psi)\approx \Omega^2(\psi)V\left(\sqrt{\frac{3}{2}}\ln\Omega(\psi)\right).
\end{equation}
In this paper, we reconstruct the potential $V_J(\psi)$ in the strong coupling limit.
The requirement that the contribution from $\omega(\psi)$ is negligible
does not necessarily mean that $\omega(\psi)\ll 1$. In fact,
we take $\omega(\psi)=1$ and $\Omega(\psi)=1+\xi f(\psi)$, where $\xi$ is the dimensionless coupling constant
and $f(\psi)$ is an arbitrary function. For simplicity, we take $f(\psi)=\psi^k$. For this specific
choice of $\Omega(\psi)$, the strong coupling conditions \eqref{strongcoup1} and \eqref{psiphirel1}
become
\begin{equation}
\label{strongcoup2}
\xi\gg \left(\frac{2}{3k^2}\right)^{k/2}\left(e^{\sqrt{2/3}\,\phi}-1\right)^{1-k}\exp\left(\sqrt{\frac{1}{6}}\,k\phi\right).
\end{equation}
The effect of the strength of the coupling constant was discussed in \cite{Broy:2016rfg}.
Note that without taking the strong coupling limit \eqref{strongcoup1}, after the conformal transformation \eqref{conftransf2},
the reconstructed potential $V_J(\psi)=\Omega^2(\psi)\, V[\sqrt{3/2}\,\ln\Omega(\psi)]$ defined in eq. \eqref{pontrel1} becomes
$V[\sqrt{3/2}\,\ln\Omega(\psi)]$ ({\it not} $V(\phi)$) in Einstein frame which depends on the coupling constant $\xi$ and the particular form of $f(\psi)$.
For different choices of the conformal factor $\Omega(\psi)$,
the forms of the potential $V[\sqrt{3/2}\,\ln\Omega(\psi)]$ are different and they depend on both $\xi$ and $f(\psi)$,
so they are different scalar-tensor theories of gravity.
In particular, in general none of them takes the reconstructed form of $V(\phi)$.
Only in the strong coupling limit, the potential $V(\sqrt{3/2}\,\ln\Omega)\approx V(\phi)$ is independent of $\xi$ and we obtain the $\xi$-attractor.
To understand the point, we review the $\xi$-attractor discussed in \cite{Kallosh:2013tua}.
In Jordan frame, we take the potentials $V_J(\psi)=\lambda^2 f^2(\psi)$ with
arbitrary functions $f(\psi)$ and the conformal factors $\Omega(\psi)=1+\xi f(\psi)$.
In Einstein frame, the potentials become
\begin{equation}
\label{veeq11}
V(\phi)=\left(\frac{\lambda f(\psi)}{1+\xi f(\psi)}\right)^2,
\end{equation}
where $\phi$ is related with $\psi$ by eq. \eqref{conftransf2}.
The potentials $V(\phi)$ in eq. \eqref{veeq11} are different for different choices of $f(\psi)$,
so the inflationary models are different.
However, in the strong coupling limit, from eq. \eqref{psiphirel1}, we get
\begin{equation}
\label{veeq12}
f(\psi)\approx \frac{e^{\sqrt{2/3}\,\phi}-1}{\xi}.
\end{equation}
Substituting eq. \eqref{veeq12} into eq. \eqref{veeq11}, in the strong coupling limit, the potentials become
\begin{equation}
\label{veeq13}
V(\phi)\approx V_0\left(1-e^{-\sqrt{2/3}\,\phi}\right)^2,
\end{equation}
where $V_0=\lambda^2/\xi^2$. So the $\xi$-attractor with $n_s=1-2/N$ and $r=12/N^2$ is reached in the strong coupling limit.

\section{The $\alpha$ attractors}

\begin{figure*}
$\begin{array}{cc}
\includegraphics[width=0.45\textwidth]{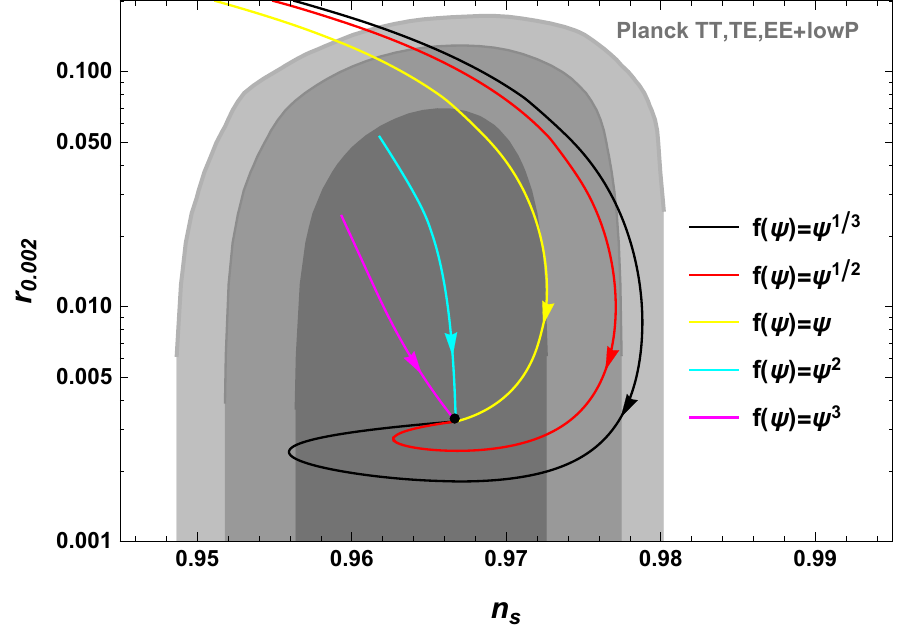}&
\includegraphics[width=0.45\textwidth]{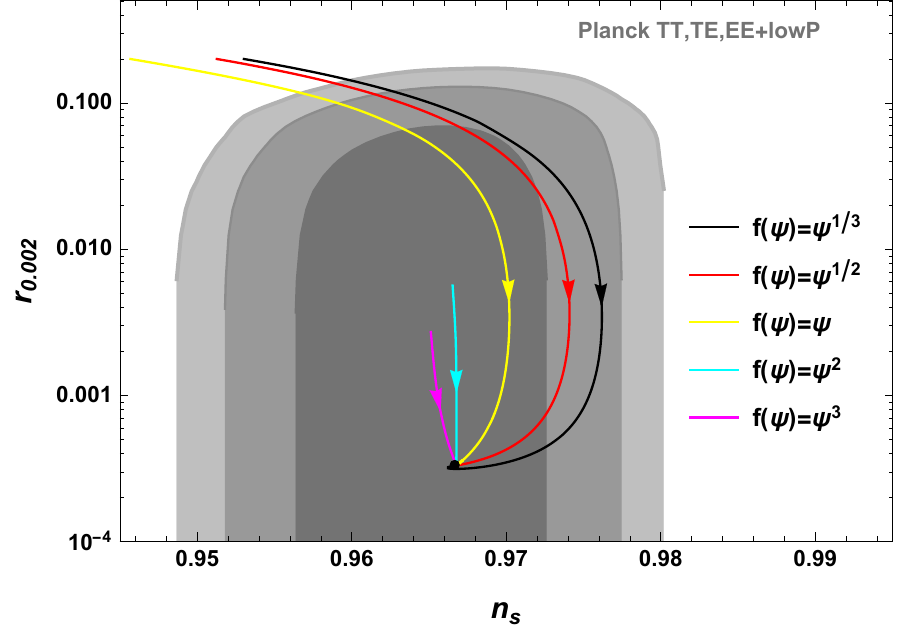}
\end{array}$
\caption{The numerical results of $n_s$ and $r$ for the scalar-tensor theory with the potential \eqref{potres4}. We take
$\Omega_0=100$, the power-law functions $f(\psi)=\psi^k$ with
$k=1/3$, 1/2, 1, 2, 3, and $N=60$. The left panel shows the $\alpha$ attractor with $\alpha=1$ and the right panel shows the $\alpha$ attractor with $\alpha=0.1$.
The coupling constant $\xi$ increases along the direction of the arrow in the plot.
The shaded regions are the marginalized 68\%, 95\% and 99.8\% C.L. contours from Planck 2015 data.}
\label{alpha}
\end{figure*}

We take the power-law parametrization
\begin{equation}
\label{rnattr1}
r=16\epsilon=\frac{12\alpha}{(N+N_0)^2}.
\end{equation}
In this paper, we assume that the parametrization is valid until the end of inflation.
At the end of inflation, $N=0$ and $\epsilon=1$, so we get $N_0=\sqrt{3\alpha/4}$.
From eq. \eqref{nsapproxeq4}, we get
\begin{equation}
\label{nsattr1}
n_s=1-\frac{2}{N+N_0}-\frac{3\alpha}{2(N+N_0)^2}\approx 1-\frac{2}{N+N_0}.
\end{equation}
This is the result for the $\alpha$ attractor \cite{Ferrara:2013rsa,Kallosh:2013yoa}.
Comparing the parametrization \eqref{rnattr1} with the Planck 2015 constraint $r<0.11$ \cite{Ade:2015lrj}, we get
$0<\alpha<35.372$ (95\% C.L.) for $N=60$. More generally, we may take the power law parametrization $r=\beta/N^p$ \cite{Mukhanov:2013tua},
then we get $n_s=1-p/N-\beta/(8N^p)$, but the observations \cite{Ade:2015lrj} tell us that $p\approx 2$,
so we consider the simple parametrization \eqref{rnattr1}.
Substituting the parametrization \eqref{rnattr1} into eqs. \eqref{nphieq2} and \eqref{sclreq2}, we get
\begin{equation}
\label{alphapont1}
V(N)=V_0\exp\left(-\frac{3\alpha}{2}\frac{1}{N+N_0}\right),
\end{equation}
\begin{equation}
\label{alphapont2}
\phi-\phi_e=\sqrt{\frac{3\alpha}{2}}\,\ln\frac{N+N_0}{N_0},
\end{equation}
where
\begin{equation}
\label{alphapont3}
\phi_e=\sqrt{\frac{3\alpha}{2}}\,\ln N_0=\sqrt{\frac{3\alpha}{8}}\,\ln \frac{3\alpha}{4}.
\end{equation}
So the reconstructed potential is \cite{Mukhanov:2013tua,Yi:2016jqr}
\begin{equation}
\label{recponteq2}
V(\phi)=V_0\exp\left[{-\frac{3\alpha}{2}\,e^{-\sqrt{\frac{2}{3\alpha}\,}\phi}}\right].
\end{equation}
When $\alpha\ll 1$, the above potential reduces to the familiar
potential $V(\phi)=V_0[1-c\exp(-\sqrt{2/3\alpha}\,\phi)]$ with $c=3\alpha/2$. Since $\phi_e<0$ when $\alpha<4/3$,
in order to reconstruct the potential $V_J(\psi)$,
we take the following transformation
in the strong coupling limit
\begin{equation}
\label{psiphirel2}
\phi\approx \sqrt{\frac{3}{2}}\ln\frac{\Omega(\psi)}{\Omega_0},\quad \Omega(\psi)\approx \Omega_0 \exp\left({\sqrt{\frac{2}{3}}\,\phi}\right),
\end{equation}
Combining
eqs. \eqref{recponteq2} and \eqref{psiphirel2}, we get the reconstructed extended inflationary potential
\begin{equation}
\label{potres4}
V_J(\psi)=V_0 \Omega^2(\psi)\exp\left[-\frac{3\alpha}{2}\left(\frac{\Omega_0}{\Omega(\psi)}\right)^{1/\sqrt{\alpha}}\right].
\end{equation}
For arbitrary function $\Omega(\psi)=1+\xi f(\psi)$, we get the same $\alpha$ attractor \eqref{rnattr1} and \eqref{nsattr1}
from the above potential \eqref{potres4} in the strong coupling limit \eqref{strongcoup1}.
If the strong coupling limit is reached, $\xi f(\psi)\gg 1$, then $\Omega(\psi)\gg 1$ and $V_J(\psi)\approx V_0\xi^2 f^2$ for $\alpha=1$. In particular,
the Higgs inflation with the nonminimal coupling function $f(\psi)=\psi^2$ is recovered.
In fig.\ref{alpha}, we take $\Omega_0=100$, $N=60$, $\alpha=1$ and $\alpha=0.1$,
$f(\psi)=\psi^{k}$ with $k=1/3$, 1/2, 1, 2 and 3 as examples to show the attractors in the strong coupling limit.
Substituting eq. \eqref{alphapont2} into eq. \eqref{strongcoup2}, in
the strong coupling limit,  the coupling constant $\xi$ satisfies
\begin{equation}
\label{strongcoup3}
\xi\gg \left(\frac{2\Omega_0}{3k^2}\right)^{k/2}\left(N+N_0\right)^{k\sqrt{\alpha}/2}
\left[\Omega_0\left(N+N_0\right)^{\sqrt{\alpha}}-1\right]^{1-k}.
\end{equation}
For $\alpha=1$, we find $\xi\gg 0.002$ with $k=3$ and $\xi\gg 1920$ with $k=1/3$.
For $\alpha=0.1$, we find $\xi\gg 0.001$ with $k=3$ and $\xi\gg 184$ with $k=1/3$.
These are confirmed by the numerical results.

From eqs. \eqref{rnattr1} and \eqref{alphapont2} and eliminating the parameter $\alpha$, we get the field excursion
\begin{equation}
\label{alplbd}
\Delta\phi=\phi_*-\phi_e=\sqrt{\frac{r}{8}}\,(N+N_0)\ln\frac{N+N_0}{N_0},
\end{equation}
which is bigger than the Lyth bound \cite{Lyth:1996im} $N\sqrt{r/8}=2\sqrt{3}\,\alpha$ as expected,
here we neglect the contribution from $N_0$.
In particular, if we take $\alpha=0.1$, then the Lyth bound gives $\Delta\phi=0.3464$.
If we use the modified Lyth bound \cite{Gao:2014pca,Gao:2014yra}, the expected field excursion would be smaller.
However, from eq. \eqref{alplbd}, we get the super-Planckian field excursion $\Delta\phi=2.0891$. As discussed in \cite{Linde:2016hbb},
the smallness of $r$ or the Lyth bound cannot be used to rule out large field inflation.

\section{Exponential parametrization}

For the simple exponential parametrization
\begin{equation}
\label{expeq1}
r=16\epsilon=C\exp(-\beta N),
\end{equation}
we get
\begin{equation}
\label{expeq2}
n_s=1-\beta-\frac{C}{8}\exp(-\beta N).
\end{equation}
Combining the parametrization \eqref{expeq1} with eqs. \eqref{nphieq2} and \eqref{sclreq2}, we derive
the potential
\begin{equation}
\label{expeq5}
V(\phi)=V_0\exp\left(-\frac{\beta}{4}\phi^2\right).
\end{equation}
The condition for the end of inflation $\epsilon(N=0)=1$ gives $C=16$.
Combining eqs. \eqref{expeq1} and \eqref{expeq2} and eliminating the parameter $\beta$, we get
\begin{equation}
\label{expeq5ab}
n_s=1+\frac{1}{N}\ln\left(\frac{r}{16}\right)-\frac{r}{8}.
\end{equation}
It is easy to show that
when $r=8/N$, $n_s$ reaches its maximum $\bar{n}_s=1-[1+\ln(2N)]/N$. If we take $N=50$, the maximum value is $\bar{n}_s=0.888$.
If we take $N=60$, the maximum value is $\bar{n}_s=0.904$. These values are
not consistent with the observations \cite{Ade:2015lrj}, so this model is excluded by the observations.

\begin{figure*}
$\begin{array}{cc}
\includegraphics[width=0.45\textwidth]{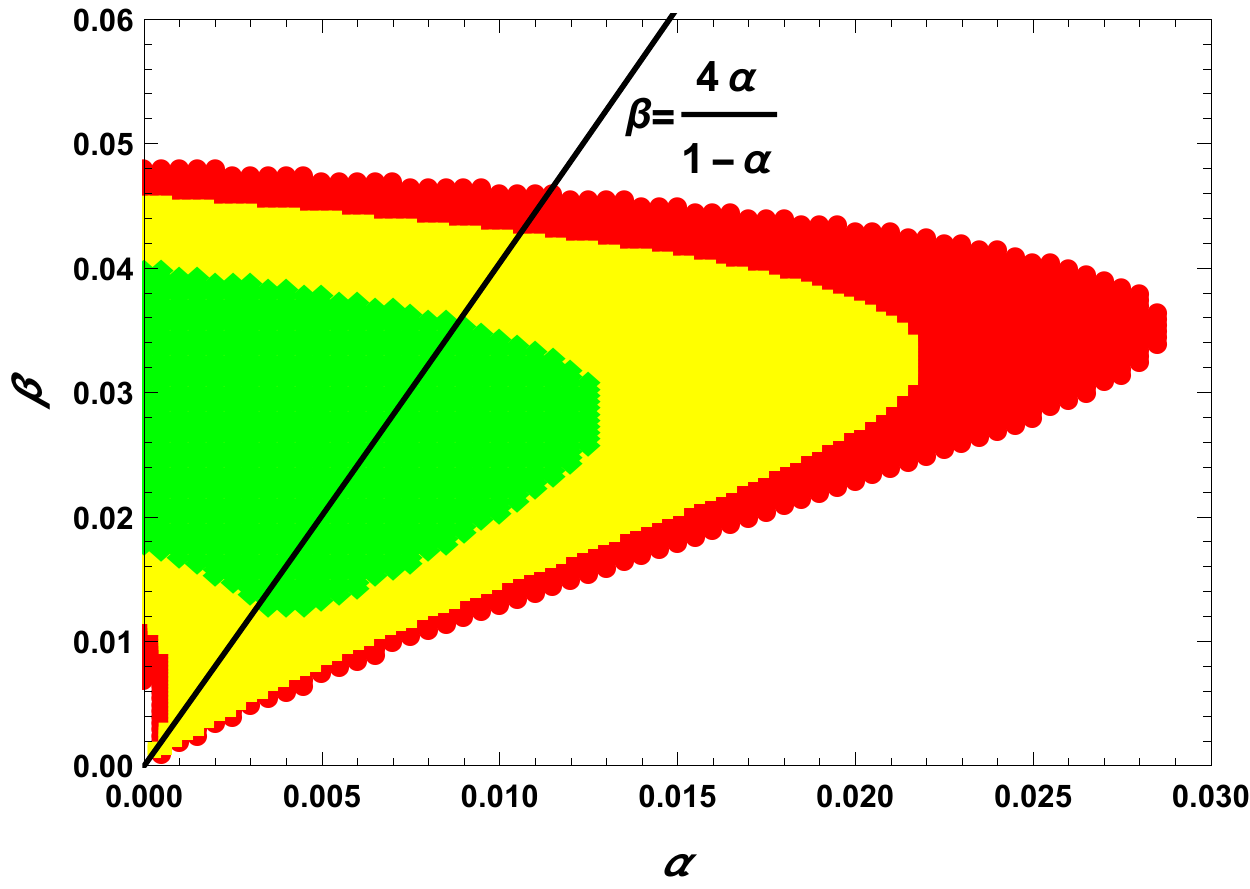}&
\includegraphics[width=0.45\textwidth]{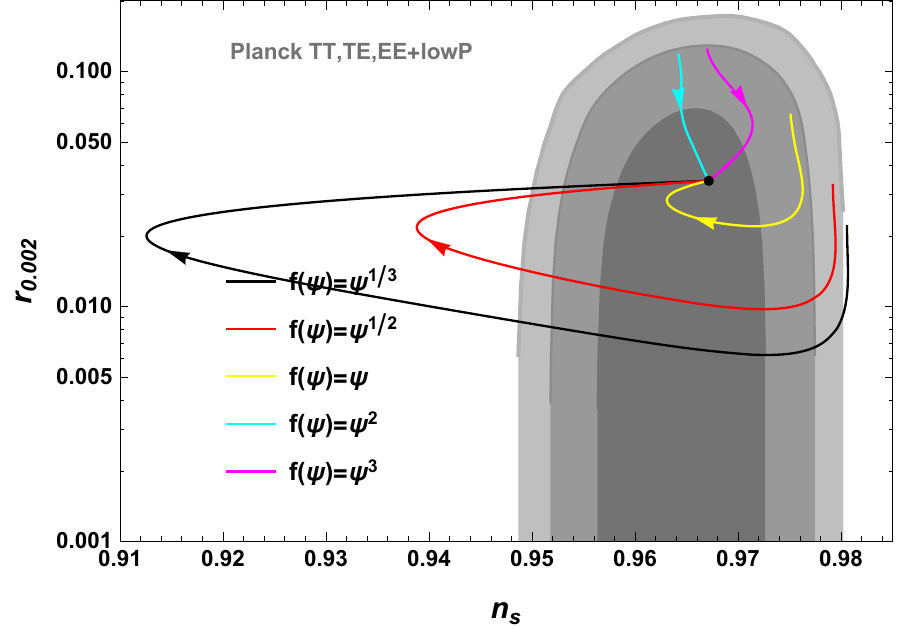}
\end{array}$
\caption{The left panel shows the constraints on $\alpha$ and $\beta$ by Planck 2015 results
for $N=60$ in the parametrization \eqref{eplneq45}.
The green, yellow and red regions correspond to 68\%, 95\% and 99.8\% C.L. contours.
The solid line $\beta=4\alpha/(1-\alpha)$ corresponds to the constant-roll inflation with the potential \eqref{eq52}.
The right panel shows the numerical results of $n_s$ and $r$ for the scalar-tensor theory with the potential \eqref{eq52}. We take
$N=60$ and the power-law functions $f(\psi)=\psi^k$ with
$k=1/3$, 1/2, 1, 2 and 3 for the constant-roll inflation with $\beta=4\alpha/(1-\alpha)$ and $\alpha=0.01/2.01$.
The coupling constant $\xi$ increases along the direction of the arrow in the plot.
The shaded regions are the marginalized 68\%, 95\% and 99.8\% C.L. contours from Planck 2015 data.}
\label{figsin}
\end{figure*}

Now we consider the parametrization \cite{Lin:2015fqa}
\begin{equation}
\label{eplneq45}
\epsilon(N)=\frac{\alpha\exp(-\beta N)}{1+s\exp(-\beta N)},
\end{equation}
where $0<\alpha<1$ and $s=-(1-\alpha)$ so that $\epsilon(N=0)=1$. The scalar spectral tilt is
\begin{equation}
\label{nsneq46}
n_s=1-\frac{\beta+2\alpha e^{-\beta N}}{1+s e^{-\beta N}}.
\end{equation}
By using the Planck 2015 results, we get the constraints on the parameters $\alpha$ and $\beta$ in the parametrization \eqref{eplneq45}
and the constraint contours are shown in the left panel of fig. \ref{figsin}.
Substituting the parametrization \eqref{eplneq45} into eq. \eqref{nsapproxeq3}, we obtain the slow-roll parameter
\begin{equation}
\label{etaeq1}
\eta=-\frac{\beta}{2}+ \frac{(4\alpha+\beta s)e^{-\beta N}}{2(1+se^{-\beta N})}.
\end{equation}
If $4\alpha+\beta s=0$, then we get the constant-roll inflationary model
with $\eta=-\beta/2$ \cite{Motohashi:2014ppa,Motohashi:2017aob}.
If the constant is 3, then it is called the utra slow-roll inflation \cite{Tsamis:2003px,Kinney:2005vj}
which is a limiting case of the constant-roll inflation.

From eq. \eqref{sclreq2}, we get
\begin{equation}
\label{eq47}
V(N)=V_0\left(1+se^{-\beta N}\right)^{-2\alpha/(\beta s)}.
\end{equation}
From eq. \eqref{nphieq2}, we get
\begin{equation}
\label{eq48}
\phi-\phi_e=\pm \frac{2\sqrt{2\alpha}}{\beta\sqrt{-s}}\left[\arccos(\sqrt{-s}\,e^{-\beta N/2})-\arccos(\sqrt{-s})\right],
\end{equation}
where
\begin{equation}\label{eq49}
\phi_e=\pm \frac{2\sqrt{2\alpha}}{\beta\sqrt{-s}}\arccos(\sqrt{-s}).
\end{equation}
Combining eqs. \eqref{eq47} and \eqref{eq48}, we get the potential
\begin{equation}\label{eq50}
V(\phi)=V_0\left[\sin\left(\frac{\beta\sqrt{1-\alpha}}{2\sqrt{2\alpha}}\phi\right)\right]^{4\alpha/[\beta(1-\alpha)]}.
\end{equation}
Choosing the superpotential $W=Xf(\Phi)$ and stabilizing the superfields $\text{Re}(\Phi)$ and $X$ at the origin, we can
obtain the inflaton potential from supergravity model as $V(\phi)=|f(\phi/\sqrt{2})|^2$ with
$\text{Im}[\Phi]=\phi/\sqrt{2}$ \cite{Kallosh:2010ug,Kallosh:2010xz,Ellis:2013xoa,Li:2013nfa,Gao:2014pca}.
Therefore, it is possible to derive the potential \eqref{eq50} from supergravity model building.

Substituting the transformation \eqref{psiphirel1} into the potential \eqref{eq50}, we get the reconstructed extended potential
\begin{equation}
\label{eq51}
V_J(\psi)=V_0\Omega^2(\psi)\left[\sin\left(\frac{\beta\sqrt{3(1-\alpha)}}{4\sqrt{\alpha}}\ln\Omega(\psi)\right)\right]^{4\alpha/[\beta(1-\alpha)]}.
\end{equation}
For arbitrary function $\Omega(\psi)$, we get the attractors \eqref{eplneq45} and \eqref{nsneq46}
in the strong coupling limit.
For the constant-roll inflation with $\eta=-\beta/2$, the potential \eqref{eq51} becomes
\begin{equation}
\label{eq52}
V_J(\psi)=V_0\Omega^2(\psi)\sin\left(\frac{\sqrt{3\beta}}{2}\ln\Omega(\psi)\right).
\end{equation}
In the right panel of fig. \ref{figsin}, we use the constant-roll inflation with $\beta=4\alpha/(1-\alpha)$ and $\alpha=0.01/2.01$ as an example to show the attractors $n_s=0.967$ and $r=0.034$ in the strong coupling limit, we take $N=60$ and $f(\psi)=\psi^{k}$ with $k=1/3$, 1/2, 1, 2 and 3.
The models with different $k$ become the model $V(\phi)=V_0\sin(0.1\phi)$ with constant-roll parameter $\eta=-0.01$ in the strong coupling limit. From eq. \eqref{strongcoup2}, we find
that the strong coupling limit requires $\xi\gg 1147$ for $k=1/3$ and $\xi\gg 0.08$ for $k=3$, these are supported by the numerical results.

Form eq. \eqref{eq48}, we get $\phi_*=9.91$.
From eq. \eqref{eq49}, we get $\phi_e=0.706$, so the field excursion is $\Delta\phi=9.21$ while the Lyth bound requires that $\Delta\phi>3.93$.

\section{Conclusions}

Under the conformal transformations \eqref{conftransf1} and \eqref{conftransf2},
in general we will get different
potential $V^i(\phi)$ in the Einstein frame from different potentials $V^i_J(\psi)$ and
arbitrary nonminimal couplings $\Omega^i(\psi)$ in the Jordan frame.
However, in the strong coupling limit \eqref{strongcoup1},
we may get the same potential $V(\phi)$ in the Einstein frame
from different potentials $V^i_J(\psi)$ with
arbitrary nonminimal couplings $\Omega^i(\psi)$ in the Jordan frame.
This is the reason for the existence of $\xi$ attractors.
In particular, the Higgs inflation with the nonminimal coupling $\xi\psi^2 R$ is a special case
of the general scalar-tensor theories of gravity with the potentials $V_J(\psi)=\lambda^2 f^2(\psi)$
and the nonminimal coupling $\xi f(\psi) R$, here the function $f(\psi)$ is an arbitrary function.
Based on this observation,
in principle we can derive any observables $n_s$ and $r$ from the general scalar-tensor
theories of gravity by choosing the potentials $V_J(\psi)$ and the nonminimal couplings $\Omega(\psi)$ appropriately.
On the other hand, we can reconstruct
inflationary potentials by parametrizing the observables such
as $r$ and $n_s$ with $N$.
The parameters in the parametrization are fitted to the observations even before we derive the potentials,
so the reconstructed models with the constrained parameters
are consistent with the observations by construction.

In this paper, we combine the idea of $\xi$ attractors and the reconstruction method to
derive the general procedure to reconstruct the extended inflationary potentials for attractors and use two
particular parametrizations \eqref{rnattr1} and \eqref{eplneq45} as examples to show how to
reconstruct the class of extended inflationary potentials in the strong coupling limit.
We show explicitly that the $\xi$ attractor is reached for different extended inflationary potentials in the strong coupling limit and
the derived results are consistent with current observations. Since in the process of reconstruction,
$\phi(N)$ can be obtained, so the field excursion is easily derived.
The result supports the conclusion that the smallness of the tensor to scalar ratio cannot rule out large field inflation.
We also derive the strong coupling condition \eqref{strongcoup2} and apply it to constrain the coupling constant $\xi$.
The derived strong coupling condition is supported by the numerical results.

In conclusion, we propose the reconstruction procedure and show that the $\xi$ attractors are reached in the strong coupling limit with
two reconstructed classes of extended inflationary potentials.
The existence of the $\xi$ attractors from extended inflationary models further challenges the model discrimination just by the observables $n_s$ and $r$.

\begin{acknowledgement}
The authors thank the center for quantum spacetime in Sogang University for the hospitality during their visits.
This research was supported in part by the National Natural Science
Foundation of China under Grant Nos. 11605061, 11875136 and 11475065,
the Major Program of the National Natural Science Foundation of China under Grant No. 11690021.
Qing Gao acknowledges the financial support from China Scholarship Council for sponsoring her visit to California Institute of Technology,
and thanks California Institute of Technology for the hospitality.
\end{acknowledgement}

%\bibliographystyle{spphys}
%\bibliography{../../book/cosmologyref}

\end{document}